\documentclass[prd,showpacs,amsmath,showkeys,twocolumn,floatfix,amssymb, preprintnumbers, nofootinbib, superscriptaddress]{revtex4} 
\usepackage{epsfig,dcolumn}
\usepackage{graphicx}
\usepackage{comment} 
\usepackage[usenames]{color}
\usepackage{graphicx}
\usepackage{bm}
 \usepackage{ifpdf}

\def\lsim{\mathrel{\rlap{\lower4pt\hbox{\hskip1pt$\sim$}}
    \raise1pt\hbox{$<$}}}
\def\gsim{\mathrel{\rlap{\lower4pt\hbox{\hskip1pt$\sim$}}
    \raise1pt\hbox{$>$}}}
\begin{document}

\title{ Coupled-channel scattering  in $1+1$ dimensional lattice model }

\author{Peng~Guo}
\email{pguo@jlab.org}
\affiliation{Thomas Jefferson National Accelerator Facility, 
Newport News, VA 23606, USA}

\preprint{JLAB-THY-13-1721 }

\date{\today}

\begin{abstract} 
 Based on the  Lippmann-Schwinger equation approach,    a generalized L\"uscher's formula in $1+1$ dimensions  for two particles scattering in both the elastic and coupled-channel cases in moving frames  is derived.  A 2D  coupled-channel scattering lattice model is presented, the  model represents a two-coupled-channel resonant scattering scalars system.  The Monte Carlo simulation is performed on finite lattices and in various moving frames. 
  The 2D   generalized L\"uscher's formula is used to extract the scattering amplitudes for the coupled-channel system from the discrete finite-volume spectrum.
  \end{abstract} 


\maketitle

\section{Introduction}
\label{intro} 

In recent years, remarkable progresses have been made on hadrons scattering in lattice QCD from both the theoretical algorithm of extracting scattering amplitudes from lattice data \cite{Lusher:1991,Gottlieb:1995,Lin:2001,Christ:2005,Bernard:2007,Bernard:2008,Liu:2005,Doring:2011,Aoki:2011,Briceno:2012yi,Hansen:2012tf,Guo:2013cp} and the practical   lattice QCD computational algorithm  aspect \cite{Michael:1985ne,Luscher:1990ck,Blossier:2009kd,Jo:2010,Edwards:2011}. 
Since  L\"uscher   proposed the elastic  scattering formalism in a finite volume \cite{Lusher:1991},  the framework has been quickly extended to   moving frames  \cite{Gottlieb:1995,Lin:2001,Christ:2005,Bernard:2007,Bernard:2008},  and to     coupled-channel scattering \cite{Liu:2005,Doring:2011,Aoki:2011,Briceno:2012yi, Hansen:2012tf,Guo:2013cp}.  The  finite volume scattering formalism  has been   successfully used by the lattice community to extract elastic hadron-hadron scattering phase shifts \cite{Aoki:2007rd, Sasaki:2008,Feng:2011,Jo_scatt:2011,Beane_scatt:2012,Lang_scatt:2011, Aoki:2011yj,Jo_scatt:2012,Jo_scatt:2013}.  Realistic lattice QCD computations on coupled-channel hadron-hadron scattering are  under way.

For the purpose of demonstrating the feasibility of  extracting coupled-channel scattering amplitudes from lattice data and   discussing some issues, such as, finite size effects,    in this work, we present a coupled-channel scattering lattice  model in 2D.    Our model is a direct generalization of  a 2D single channel scattering lattice model in \cite{Gatteringer:1993}. The advantage of scattering in 2D is that   only finite numbers of scattering amplitudes  contribute  in one spatial dimensional scattering theory, and the relation between phase shift and energy level  in  L\"uscher's  formula in 2D \cite{Luscher:1990ck} appears more transparent. Our 2D lattice model represents a   coupled-channel  resonant scattering system with three species of scalar fields $(\phi,\sigma, \rho)$, where the scalar field $\rho$ acts as a resonance which couples to both $2\phi$ and $2\sigma$ channels. The Monte Carlo simulation are carried out  on various  lattice sizes and in different moving frames. We also present the derivation of 2D L\"uscher's  formulae in a general moving frame and for a coupled-channel system. The derivation is  based on the Lippmann-Schwinger equation approach  presented in \cite{Guo:2013cp}. These formulae are used in the end to extract the scattering amplitudes from Monte Carlo  simulation data. The finite size effect on extracting scattering amplitudes (phase shifts and inelasticity) from lattice data is also addressed in this work. Although, our   model is  formulated and computed in $1+1$ dimensions, it still captures  many of the features  of hadrons scattering in a real $3+1$ dimensional  QCD computation,  and sheds some light on the future  coupled-channel hadron-hadron scattering lattice QCD calculation.

The paper is organized as follows. A discussion of elastic scattering in a finite volume is given   in Section \ref{singlechannel}, with extension to the coupled channel system in Section \ref{coupledchannel}.  The 2D lattice model, the Monte Carlo simulation  and data analysis   are described in Section \ref{isingmodel}. The summary and outlook are given in Section \ref{summary}.

\section{ Lu\"scher's formula in $1+1$ dimensions }\label{singlechannel}

For completeness, we first present the basic scattering theory in $1+1$ dimensions. Based on  the  Lippmann-Schwinger equation approach,     a generalized L\"uscher's formula in $1+1$ dimensions  for two particles elastic scattering   in moving frames is presented   in the end of this  Section.

\subsection{Two-particle scattering in infinite volume}
\begin{figure}
\begin{center}
\includegraphics[width=2.6 in,angle=0]{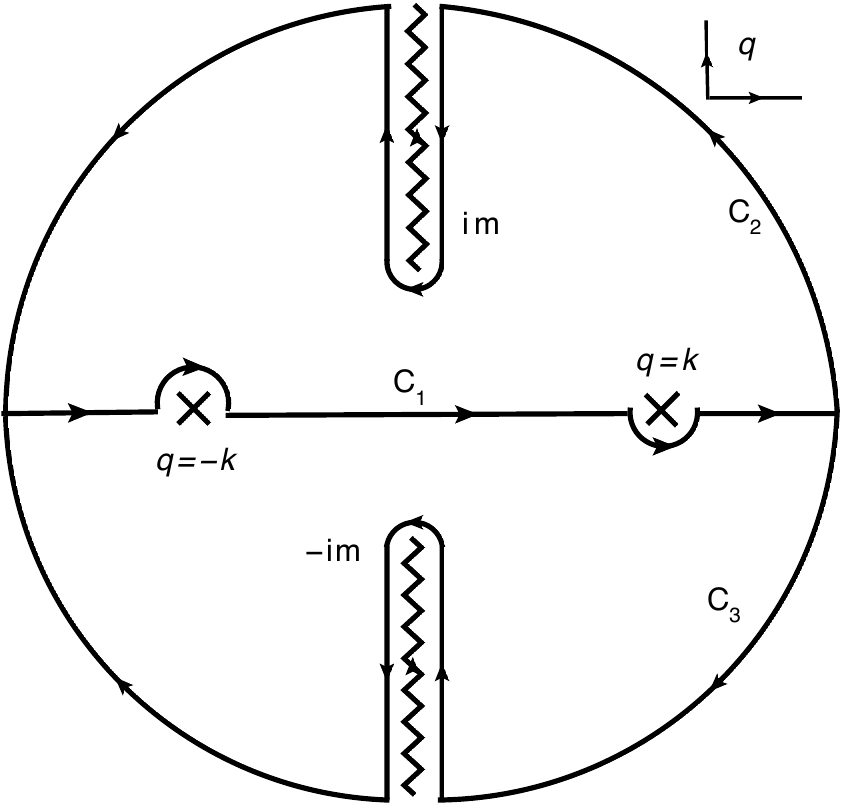}  
\caption{The integration contours and singularities of free Green's function in Eq.(\ref{green_free}) on complex $q$ plane.
\label{green_cut}}
\end{center}
\end{figure}

We consider spinless particles  scattering in    a symmetric potential   $\tilde{V}(-x) = \tilde{V}(x) $, the mass of scalar particles is $m$.  The wave function of  scattering   particles   in center of mass frame   satisfies   the relativistic Lippmann-Schwinger  equation
\begin{equation}\label{lippmann}
 \psi(x) =\int_{-\infty}^{\infty} dx' G_{0}(x-x'; \sqrt{s})  \tilde{V}(x')  \psi(x') ,
\end{equation}
where  the center of mass frame energy is $\sqrt{s}$ and the free-particle Green's function is given by
\begin{equation}\label{green_free}
G_{0}(x; \sqrt{s})  = \int_{-\infty}^{\infty} \frac{d q}{2\pi} \frac{e^{ i q x}}{ \sqrt{s}-2 \sqrt{q^{2}+m^{2}}}.
\end{equation}
The Green's function can be further written as a oscillating term and an exponentially decaying term over the separation of two particles. The singularities of integrand in Eq.(\ref{green_free}) on  the complex $q$ plane are    two poles  on real axis $q=\pm k$ and two branch cuts   on imaginary axis   $\pm \left [im , i \infty \right ]$, see Fig.\ref{green_cut}. Therefore,  for $x>0$, we choose  the contour  $C_{1}+C_{2}$ to include pole $q=k$ and cross the cut $\left  [im , i \infty \right ]$,  and  for $x<0$, we  choose  the contour $C_{1}+C_{3}$ to include pole  $q=-k$ and cross the cut  $- \left [im , i \infty \right ]$, as shown in  Fig.\ref{green_cut}.  Thus, contour integral leads to
         \begin{equation}\label{contour_green}
G_{0}(x  ,\sqrt{s}) = - i  \frac{\sqrt{s}}{4 k}  e^{i k  |x| } - \int_{m }^{ \infty }   \frac{  d \rho }{2\pi }      \sqrt{   \rho^{2}-m^{2}}       \frac{ e^{ -\rho |x|}  }{   k^{2}+\rho^{2}     }     ,   
   \end{equation}   
 where $k = \frac{\sqrt{s- 4 m^{2}}}{2}$ is  momentum of   particle in CM frame.  At large separations, the free Green's function can be approximated by  the oscillating term only
 \begin{align} 
G_0& (x-x'  , \sqrt{s})   \nonumber \\ & \stackrel{ |x|>|x'| }{  \simeq }    - i  \frac{\sqrt{s}}{4 k}  e^{i k  |x| }  \sum_{\mathcal{P}= \pm } Y_{\mathcal{P}} (x) Y_{\mathcal{P}} (x')  J^{*}_{\mathcal{P}}(k x'),  
\end{align}
     where  the functions $Y_{\mathcal{P}} (x)$ and $ J_{\mathcal{P}}(k x)$ are defined by
     \begin{eqnarray}
   &&  Y_{+}(x)=1, \ \  Y_{-}(x) = \frac{x}{|x|}, \\
    && J_{+}(k x) = \cos k |x|, \ \ J_{-} (kx) = i \sin k|x|.
     \end{eqnarray}
Such that   $Y_{\mathcal{P}} (x)$ and $ J_{\mathcal{P}}(k x)$  resemble the spherical harmonic and Bessel functions in three spatial dimensions, and  $Y_{\mathcal{P}}(x)$  is the parity  eigenstate with eigenvalue $\mathcal{P}$.  The continuous rotation symmetry in three dimensions reduces to discrete spatial reflection $x \rightarrow -x$  in one spatial dimension, thus,   the partial wave expansion of wave function in three dimensions reduce to the expansion of  the wave function in terms of   parity eigenstates  $\psi(x) = \sum_{\mathcal{P}=\pm} c_{\mathcal{P}} \psi_{\mathcal{P}}(x)$, where $  \psi_{\mathcal{P}}(-x) =\mathcal{P}  \psi_{\mathcal{P}}(x)$. 

For a potential $\tilde{V}$ which falls at large separations,   Eq.(\ref{lippmann})  is solved outside the range of the potential by
\begin{equation}\label{wavefree}
\psi(x)  \stackrel{ |x|>R }{  \longrightarrow } \sum_{\mathcal{P} =\pm} c_{\mathcal{P}} Y_{\mathcal{P}}(x) \left [J_{\mathcal{P}} (kx)+ i e^{i k |x|}  f_{\mathcal{P}}(k) \right ] ,
\end{equation}
   where $R$ denotes to the effective range of potential and the free solutions has been also included in Eq.(\ref{wavefree}). The scattering amplitudes are defined by
   \begin{equation}\label{ampsingle}
 c_{\mathcal{P}}  f_{\mathcal{P}}(k) = - \frac{\sqrt{s}}{4 k} \int_{-\infty}^{\infty} d x'  Y_{\mathcal{P}}(x') J^{*}_{\mathcal{P}} (kx') \tilde{V}(x') \psi(x'),
   \end{equation}
   which up to the inelastic threshold can be parametrized by scattering phase shift
   \begin{equation}
   f_{\mathcal{P}}(k) = e^{i\delta_{\mathcal{P}} } \sin \delta_{\mathcal{P}}.
   \end{equation}

\subsection{Two-particle scattering on  a torus}
Now we consider the theory in a one spatial  dimensional box with periodic boundary conditions. In lattice QCD calculations,  the computations are usually done in the moving frame of the two-particle system \cite{Gottlieb:1995}. After  the system is boosted back to the CM frame, the  shape of cubic  box in moving frame is deformed in CM frame due to Lorentz contraction.  Similarly, in the one spatial dimension,  the volume of a one dimensional box, $L$, in a moving frame with total momentum $P= \frac{2\pi}{L} d, d\in \mathbb{Z}$  becomes   $\gamma L$ in CM frame,  where $\gamma =\sqrt{1+ \frac{P^{2}}{s}}$ is the Lorentz contraction factor. 

 Taking into account the Lorentz contraction effect as well, we divide the integral over $x'$ into a sum of integrals over each translated box in Eq.(\ref{lippmann}), giving, 
 \begin{align} 
\psi^{(L)}(x) &= \sum_{n \in \mathbb{Z}} \int_{-\frac{ \gamma L}{2}}^{\frac{ \gamma L}{2}} dx' G_{0}(x-x' - \gamma n L; \sqrt{s})   \nonumber \\ &  \times  \tilde{V}(x'+ \gamma nL)  \psi^{(L)}(x' + \gamma nL).
\end{align}
The wave function in CM frame satisfies the boundary condition \cite{Gottlieb:1995}  of  
\begin{equation} 
   \psi^{(L)}(x + \gamma nL)  = e^{i \frac{P }{2} n L} \psi^{(L)}(x),
\end{equation}
   Using the periodicity of the potential $ \tilde{V}(x'+ \gamma nL) =  \tilde{V}(x')$, we have
\begin{equation} 
  \psi^{(L,P)}(x) =  \int_{-\frac{\gamma L}{2}}^{\frac{\gamma L}{2}} dx' G_{P}(x-x' ; \sqrt{s})   \tilde{V}(x')  \psi^{(L,P)}(x' ) , 
\end{equation}
where the periodic Green's function is given by
\begin{equation}\label{greensum} 
 G_{P}(x-x' ; \sqrt{s})   = \sum_{n \in \mathbb{Z}}  G_{0}(x-x' - \gamma  n L; \sqrt{s})   e^{i \frac{P}{2} n L}.
\end{equation}
By using the Poisson summation formula, $\frac{1}{2\pi} \sum_{n\in\mathbb{Z}}  e^{i \frac{P}{2} n L} = \frac{1}{ \gamma L} \sum_{n \in \mathbb{Z}} \delta ( \frac{P}{2 \gamma}+ \frac{2\pi}{ \gamma L} n)$, Eq.(\ref{greensum}) can be reexpressed as
\begin{equation}\label{greenperiodic} 
 G_{P}(x-x' ; \sqrt{s})   = \frac{1}{\gamma L}  \sum_{ q \in P_{d}}      \frac{e^{ i q ( x-x')}}{ \sqrt{s}-2 \sqrt{q^{2}+m^{2}}},
\end{equation}
where $P_{d}=\{ q \in \mathbb{R} | q = \frac{2\pi}{ \gamma L} (n + \frac{d}{2} ) , \mbox{for} \  n \in \mathbb{Z} \}$.

As in the infinite volume case, the periodic Green's function Eq.(\ref{greenperiodic}) can be shown to consist of an oscillatory part and an exponentially decaying part which can be neglected for large volume $\gamma L> m^{-1}$. The remaining oscillatory part takes the form
\begin{equation}\label{greenpole} 
 G_{P}(x-x' ; \sqrt{s})    \rightarrow - i \frac{\sqrt{s}}{4k} \sum_{n \in \mathbb{Z}}  e^{ i k |x  -x'- \gamma n L|} e^{i \frac{P}{2} n L},
\end{equation}
where  we have used    Eq.(\ref{contour_green}) and Eq.(\ref{greensum}). The infinite sum in Eq.(\ref{greenpole}) can be done analytically,  the details are presented in Appendix \ref{expgreen}, so that
\begin{align} 
G_{P}&(x-x' ; \sqrt{s})  \nonumber \\  & \stackrel{ |x|>|x'| }{  \simeq }   - i \frac{\sqrt{s}}{4k}  \sum_{\mathcal{P} =\pm} Y_{\mathcal{P}} (x) Y_{\mathcal{P}} (x')  J^{*}_{\mathcal{P}} (kx')    \nonumber \\ & \quad\quad    \times \left [ e^{ i k |x  |} - \left (1- i \cot \frac{\gamma k L + \pi d}{2} \right )    J_{\mathcal{P}} (k x) \right ]. 
\end{align}
Using the definition of scattering amplitudes in Eq.(\ref{ampsingle}), we can express the wave function as
\begin{align}\label{wavefinite} 
  \psi^{(L,P)}  (x) &  \stackrel{ |x|>R }{  \longrightarrow }    \sum_{\mathcal{P} = \pm}  c_{\mathcal{P}} Y_{\mathcal{P}} (x)  f_{\mathcal{P}} (k)  \nonumber \\
&    \quad   \times \left [  i e^{ i k |x  |} -  \left (i+ \cot \frac{\gamma k L + \pi d}{2} \right )    J_{\mathcal{P}} (k x) \right ].
\end{align}

Matching the wave function in finite box given by Eq.(\ref{wavefinite}) to the wave function in infinite volume given by Eq.(\ref{wavefree}) at a arbitrary $|x| >R$, we obtain
\begin{align} 
  \sum_{\mathcal{P} = \pm} & c_{\mathcal{P}} Y_{\mathcal{P}}(x) J_{\mathcal{P}} (kx) f_{\mathcal{P}} (k)  \nonumber \\
 &\times \left [ \frac{1}{f_{\mathcal{P}} (k)} +      i+ \cot \frac{\gamma k L + \pi d}{2} \right  ]=0 .
\end{align}
which has non-trivial solution when
\begin{equation}\label{singeq} 
\cot \delta_{\mathcal{P}}+ \cot \frac{\gamma k L + \pi d}{2}   =0 .
\end{equation}

\section{ Coupled-channel scattering in $1+1$ dimensions }\label{coupledchannel}  
The previous discussion in section \ref{singlechannel} can be generalized to a coupled channel system by including another species of scalar fields, let's name two species of particles $\phi$ and $\sigma$, the masses  are  $m_{\phi ,\sigma}$.   The coupled channel wave function  has the form of  $\psi(x) = \sum_{\alpha = \phi, \sigma  }  \psi^{\alpha} (x) $, and $ \psi^{\alpha} (x) $ satisfies equation
\begin{equation} 
 \psi^{\alpha}(x) =\int_{-\infty}^{\infty} dx' G^{\alpha}_{0}(x-x'; \sqrt{s})  \sum_{\beta= \phi, \sigma} \tilde{V}_{\alpha \beta}(x')  \psi^{\beta}(x') ,  
\end{equation}
A $2\times 2$ matrix of coupled-channel scattering amplitudes can be defined by
\begin{equation}
c^{\alpha}_{\mathcal{P}}  f^{\alpha \beta}_{\mathcal{P}} = - \frac{\sqrt{s}}{4 k_{\alpha}} \int_{-\infty}^{\infty} d x'  Y_{\mathcal{P}}(x') J^{*}_{\mathcal{P}} (k_{\alpha}x') \tilde{V}_{\alpha \beta}(x') \psi^{\beta}(x'),  
\end{equation}
where $k_{\alpha} = \frac{\sqrt{s-4 m_{\alpha}^{2}}}{2}$ is the CM frame scattering momentum in channel $\alpha$. Neglecting exponentially decaying terms and also include the free solution, we have for the wave function in channel $\alpha$
\begin{align}\label{coupwavefree}
 \psi^{\alpha}(x) & \stackrel{ |x|>R }{  \longrightarrow }  \sum_{\mathcal{P} =\pm} Y_{\mathcal{P}}(x) \nonumber \\
& \quad \times \left [c^{\alpha}_{\mathcal{P}}  J_{\mathcal{P}} (k_{\alpha}x)+ i e^{i k_{\alpha} |x|}  \sum_{\beta} c^{\beta}_{\mathcal{P}} f^{\alpha \beta}_{\mathcal{P}} \right ] .  
\end{align}

Extending the single channel derivation in finite-volume to the two-channel system, one obtains,
\begin{align}\label{coupwavefinite} 
  \psi^{\alpha(L,P)}(x)  & \stackrel{ |x|>R }{  \longrightarrow }     \sum_{\mathcal{P} =\pm}  \sum_{\beta}  c^{\beta}_{\mathcal{P}} Y_{\mathcal{P}} (x)  f^{\alpha \beta}_{\mathcal{P}}  \nonumber \\
&   \times \left [  i e^{ i k_{\alpha} |x  |} - \left (i+ \cot \frac{\gamma k_{\alpha} L + \pi d}{2} \right )    J_{\mathcal{P}} (k_{\alpha} x)  \right ].
\end{align}
Matching the wave function in finite-volume, Eq.(\ref{coupwavefinite}) to the wave function in infinite volume, Eq.(\ref{coupwavefree}), we can derive a condition for non-trivial solutions
\begin{widetext}
\begin{equation}\label{coupcondition}
\left (\frac{1}{f^{\phi \phi}_{\mathcal{P}}} +i + \cot \frac{\gamma k_{\phi} L + \pi d}{2} \right ) \left (\frac{1}{f^{\sigma \sigma}_{\mathcal{P}}} +i + \cot \frac{\gamma k_{\sigma} L + \pi d}{2}  \right )= \left (i + \cot \frac{\gamma k_{\phi} L + \pi d}{2} \right ) \left (i + \cot \frac{\gamma k_{\sigma} L + \pi d}{2} \right ) \frac{ \left (f^{\phi \sigma}_{\mathcal{P}} \right  )^{2}}{f^{\phi \phi}_{\mathcal{P}} f^{\sigma \sigma}_{\mathcal{P}}} .   
\end{equation}
\end{widetext}
The scattering amplitudes can be parametrized by three real parameters: two phase shifts $\delta^{\alpha}_{\mathcal{P}}$ and an inelasticity $\eta_{\mathcal{P}}$,
\begin{equation}\label{scattamp}
 f^{\alpha \alpha}_{\mathcal{P}} = \frac{\eta_{\mathcal{P}} e^{ 2 i \delta_{\mathcal{P} }^{\alpha  } }-1}{2 i },   f^{\alpha \beta}_{\mathcal{P}} = \frac{ \sqrt{1- \eta^{2}_{\mathcal{P}} } e^{  i  \left ( \delta_{\mathcal{P}  }^{\alpha  }  + \delta_{\mathcal{P}  }^{  \beta} \right ) }   }{2  }.
\end{equation}
Thus, we can also write the Eq.(\ref{coupcondition})  as
 \begin{equation}\label{coupeq}  
\eta_{\mathcal{P}} \left (-1 \right )^{d} = \frac{\cos \left (\gamma L \frac{ k_{\phi} + k_{\sigma}}{2} + \delta^{\phi}_{\mathcal{P}} + \delta^{\sigma}_{\mathcal{P} } \right )}{\cos \left (\gamma L \frac{ k_{\phi} - k_{\sigma}}{2} + \delta^{\phi}_{\mathcal{P}} - \delta^{\sigma}_{\mathcal{P} } \right)}.
\end{equation}

\section{ The Ising model for coupled channel scattering}\label{isingmodel}
To simulate  a coupled channel scattering system,  we build a model with two light mass fields $(\phi, \sigma)$ coupled to a heavier mass field $\rho$ with two 3-point couplings,  $\rho \phi^{2}$ and $\rho \sigma^{2}$. The physical masses of the fields are calibrated to be at the region   $2m_{\phi}<2 m_{\sigma}<m_{\rho}< 4 m_{\phi}$. For elastic scattering, the Ising model has been used and tested in both $1+1$ \cite{Gatteringer:1993} and $3+1$ \cite{Gottlieb:1995} dimensions by coupling two Ising fields, $\phi $ and $\rho$, together through a 3-point nonlocal interaction. For our purpose, we could introduce one more species of Ising field $\sigma$ and another 3-point  term to couple $\sigma$ and $\rho$ together, where $\rho$ field gives rise to the resonant  behavior in both $\phi \phi$ and $\sigma \sigma$ channels.

 The action is given by
\begin{align}\label{action}
S=& - \sum_{\alpha = \phi, \sigma, \rho} \kappa_{\alpha} \sum_{x, \mu} \alpha(x) \alpha(x+\hat{\mu})  \nonumber \\
&+ \sum_{\beta = \phi, \sigma} g_{\rho \beta \beta} \sum_{x, \mu} \rho(x) \beta(x) \beta(x+\hat{\mu}),
\end{align}
where $x=(x_{0},x_{1})$ is coordinates of Euclidean $T\times L$ lattice site and $\hat{\mu}$ denotes the unit vector in direction $\mu$. The values of the fields are restricted to $\pm 1$, and the periodic boundary condition has been applied in Monte Carlo simulation. In the scaling limit, the Ising model represent a lattice $\phi^{4}$ theory, thus,   the action in Eq.(\ref{action})  effectively  describes an interacting theory of \cite{Gottlieb:1995} 
\begin{align}
S=&\sum_{\alpha = \phi,\sigma, \rho}  \int d^{2}x \left [\frac{1}{2} \left (\partial \alpha \right )^{2} + \frac{1}{2} m_{\alpha}^{2} \alpha^{2} + \frac{\lambda_{\alpha}}{4!} \alpha^{4} \right ] \nonumber \\
&+ \int d^{2} x  \left(  \frac{g_{\rho \phi \phi}}{2} \rho \phi^{2} + \frac{g_{\rho \sigma \sigma}}{2} \rho \sigma^{2} \right )
 \end{align}
in Euclidean space.  By adjusting the masses and coupling constants,  we could have  an resonance $\rho$ sit above both $2 \phi$ and $2 \sigma$ thresholds, and couple to both channels by interaction terms   $g_{\rho \phi \phi} \rho \phi^{2}$ and $g_{\rho \sigma \sigma} \rho \sigma^{2}$ respectively. Therefore, the lattice Monte Carlo simulation by using the action in Eq.(\ref{action}) is expected to imitate a coupled-channel scattering model: $\phi \phi +\sigma \sigma  \leftrightarrow \rho \leftrightarrow \phi \phi +\sigma \sigma $. 
  Due to the Bose-symmetry,    only scattering amplitudes   with   positive parity contribute   in this model.

\subsection{Cluster algorithm for coupled-channel Ising model}
An generalized cluster algorithm is used in our simulation, similar to the cluster algorithm in  \cite{Gatteringer:1993}, we update $\rho$, $\phi$ and $\sigma$ fields alternately.

Updating the $\rho$ field: Bonds between neighbored spins of equal sign are kept with the probability  $1-e^{-2 \kappa_{\rho}}$.  After identification of the connected clusters, the spin of cluster is flipped with probability
\begin{eqnarray}
&&p_{\rho}^{\text{flip}} = \frac{1}{1+e^{-2 \alpha (C)}}, \\
&&\alpha(C) = \sum_{\beta = \phi, \sigma} g_{\rho \beta \beta} \sum_{x \in C, \mu} \rho(x) \beta(x) \beta(x+\hat{\mu}). 
\end{eqnarray}

Updating the $ \beta =\phi,\sigma$ fields: Bonds between like-sign neighbors are kept with the probability  $1-e^{-2 \left [\kappa_{\beta} -g_{\rho \beta \beta}  \frac{\rho(x) + \rho(x+\hat{\mu})}{2} \right  ]}$, the spin of cluster is flipped with probability $\frac{1}{2}$.

In our simulation, the parameters are chosen  as $\kappa_{\rho}=0.3323, \kappa_{\rho}=0.3897, \kappa_{\sigma} =0.3748$, and $g_{\rho \phi \phi} = g_{\rho \sigma \sigma} = 0.02$, the masses of $\phi$ and $\sigma$ fields are measured through single particle propagators, the values are given by $m_{\phi} \simeq 0.176$ and $m_{\sigma} \simeq 0.240$ respectively  in lattice unit. The mass of resonance $\rho$ is established from phase shifts, and the approximate value is given by  $m_{\rho} \simeq 0.57$.

In this work,  we use $T=80$  and various spatial extensions $L$ between 15 and 50. For each set of lattice size  and moving frame, we generated  typically one million measurements.

\subsection{Particles spectrum}
\begin{figure}
\begin{center}
\includegraphics[width=0.54\textwidth]{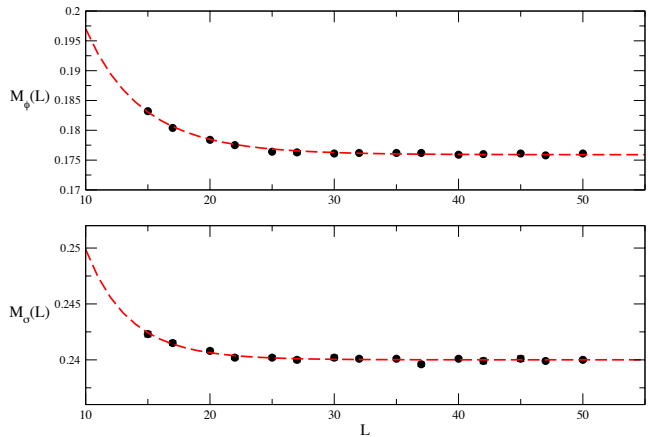}  
\caption{ $m_{\phi, \sigma}$ as function of $L$,  they follow the function of $M_{\alpha}(L) = m_{\alpha} + c_{\alpha}/L^{1/2}e^{ - m_{\alpha} L}$ (red dashed curves). \label{singlemass}}
\end{center}
\end{figure}

As shown in the elastic scattering case in $1+1$ dimensions  \cite{Gatteringer:1993},  one particle propagator can be constructed by operators 
\begin{equation}
\tilde{\alpha}_{n}(x_{0}) =\frac{1}{L} \sum_{x_{1}} \alpha(x) e^{i x_{1} q_{1, n}}, 
\end{equation}
where $q_{1, n} = \frac{2\pi}{L} n, n = - L/2 +1, \cdots, L/2$ and $\alpha = \phi, \sigma$. The spectrum of single particle fields is extracted from exponential decay of the correlation functions
\begin{equation}
C_{\alpha, n}(x_{0}) = \langle   \tilde{\alpha}_{-n}(x_{0}) \tilde{\alpha}_{n}(0) \rangle \propto e^{-E^{\alpha}_{q} x_{0}}.
\end{equation}
 The single particle's masses satisfy relation $M_{\alpha}(L) = m_{\alpha} + c_{\alpha}L^{-1/2} e^{- m_{\alpha } L }$ \cite{Gatteringer:1993}, see Fig.\ref{singlemass}.

The two particles operators in the moving frame with total momentum of $P= \frac{2\pi}{L} d , d\in \mathbb{Z}$ are constructed  from single particle operators
\begin{eqnarray}
&&O^{d}_{(\rho,d)}(x_{0}) = \tilde{\rho}_{d}(x_{0}) , \\
&& O^{d}_{(\alpha,n)}(x_{0}) = \tilde{\alpha}_{n}(x_{0})   \tilde{\alpha}_{d-n}(x_{0}) .
\end{eqnarray}
The  two particles correlation function matrices read
\begin{equation}
C^{d}_{i j} (x_{0}) = \langle  \left [  O^{d*}_{i}(x_{0}) -  \delta_{d,0} O^{d*}_{i}(x_{0}+1)   \right ] O^{d}_{j}(0)    \rangle ,
\end{equation}
where  short hand notation  $i,j $ denotes the different  sets of  $ (\rho,d)$ or $ (\alpha, n)$. The disconnected contribution has to be subtracted in CM frame ($d=0$). The spectral decomposition of the correlation function matrices has the form, 
\begin{equation}
C^{d}_{i j} (x_{0}) = \sum_{l} v^{(d,l) *}_{i}v^{(d,l) }_{j} e^{- E_{l}^{(d)} x_{0}},
\end{equation}
where $v^{(d,l) }_{i} =\langle l  |O^{d}_{i}(0) |0\rangle $ and $l$ labels the energy eigenstate $E_{l}^{(d)}$. The energy levels are determined by solving generalized eigenvalue problem  \cite{Luscher:1990ck}   
\begin{equation}
C^{d}(x_{0}) \xi_{l} = \lambda_{(d,l)} (x_{0}, \bar{x}_{0}) C^{d}(\bar{x}_{0}) \xi_{l} ,
\end{equation}
where $\lambda_{(d,l)} (x_{0}, \bar{x}_{0}) = e^{- (x_{0} -\bar{x}_{0}) E_{l}^{(d)}} $ and   $\bar{x}_{0}$ is a small reference time, in our analysis, $\bar{x}_{0}$ is set to be zero. In our simulation,  the size of the matrices varies according to the volume, the number of operators we are using is always two or three more than the number of energy eigenstates in the region $2 m_{\phi} < \sqrt{s} < 4 m_{\phi}$. The values of the energy levels are determined by fitting $\lambda_{(d,l)} (x_{0},0)  $ for $0 \leqslant x_{0} \leqslant  6-10$  with the form 
\begin{equation}
\lambda_{(d,l)}  (x_{0},0)  =\left (1- A_{(d,l)}  \right ) e^{- m_{(d,l)}  x_{0}} + A_{(d,l)} e^{- m'_{(d,l)}  x_{0}},  \nonumber  
\end{equation}
where $ A_{(d,l)} , m_{(d,l)}  $ and $m'_{(d,l)}  $ are fitting parameters. This form allows a second exponential, however, we found that the  value of $m'_{(d,l)}  $ is typically $2-3$ times  of  the value of $m_{(d,l)}  $, so that it decreases rapidly and the first exponential becomes  dominant at around $x_{0}=2$.

\begin{figure}
\begin{center}
\includegraphics[width=0.54\textwidth]{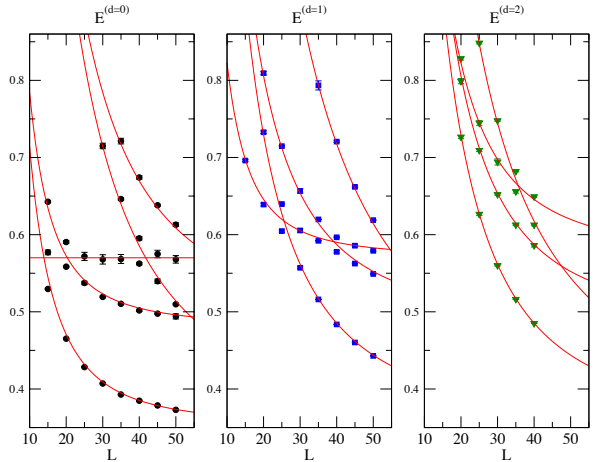}  
\caption{ The  energy spectra of a coupled-channel Ising model as function of $(L,d)$: (Left) $d=0$, (Middle) $d=1$ and (Right) $d=2$. The red curves represent (I) the energy spectra of a non-interacting pair of particles:  $E^{(d)} = \sum_{i=\pm} \cosh^{-1} \left ( \cosh m_{\phi, \sigma} +1 -\cos  p_{i}  \right )$, where $p_{\pm }=\frac{2 \pi}{L} n_{\pm} \pm \frac{\pi}{L}$, $n_{+}+n_{-} = d$,  and  $(n_{\pm},  d) \in \mathbb{Z}$. The masses of $\phi, \sigma$ are given by $m_{\phi} \simeq 0.176, m_{\sigma} \simeq 0.240$; (II) the    energy spectra of a stable resonance: $E^{(d)} =  \cosh^{-1} \left ( \cosh m_{\rho} +1 -\cos  P  \right )$, where $m_{\rho} \simeq 0.57$.  \label{spec}}
\end{center}
\end{figure}

We show the measured two-particle energy spectra from our simulations in Fig.\ref{spec} for various volumes and   total momenta of  two particles system $P=\frac{2\pi}{L}d, d=0,1,2$.

\subsection{A coupled-channel $K$-matrix model}\label{kmat}

In order to extract the  scattering amplitudes (phase shifts and inelasticity) from the discrete finite volume spectra of the Monte Carlo simulation, we   consider a   $K$-matrix model for a coupled-channel S-wave scattering system.

In the scaling regime, the phase shifts of  the Ising model $\delta_{\phi, \sigma}$ in $1+1$ dimensions  are shifted by a background phase $\delta_{\text{Ising}} = \frac{\pi}{2} $ \cite{Gatteringer:1993, Sato:1977, Berg:1978}, 
\begin{equation}
\delta_{\phi, \sigma} = \delta^{\text{Res}}_{\phi, \sigma} -\delta_{\text{Ising}} , 
\end{equation} 
where $ \delta^{\text{Res}}  $  represent the normal phase shift in which  a resonance may appear at value of $ \delta^{\text{Res}}= \frac{\pi}{2}$. Thus, the unitarized   $t$-matrix may be defined by
\begin{eqnarray}
t_{\alpha \alpha} &=& - t^{\text{Res}}_{\alpha \alpha} +i  \theta(s- 4 m_{\alpha}^{2} ) \frac{ \sqrt{s}}{2 k_{\alpha}}, \nonumber \\
 t_{\alpha \beta} &=& - \theta(s- 4 m_{\sigma}^{2} )  t^{\text{Res}}_{\alpha \beta},  \nonumber
\end{eqnarray}
where $t-$ and $t^{\text{Res}}$-matrix  are parametrized by   phase shifts $\delta_{\phi, \sigma}, \ \delta^{\text{Res}}_{\phi, \sigma}$ and    inelasticity $\eta$ respectively.  The $t$-matrix is related to the$f$-matrix defined   in Eq.(\ref{scattamp}) by equation $t_{\alpha \beta}   = \frac{\sqrt{s}}{2 \sqrt{k_{\alpha} k_{\beta}}} f_{\alpha \beta} $. The unitarity of  the
$t$-matrix is guaranteed by the unitarity of  the      $t^{\text{Res}}$-matrix.  The  Ising model suggestion is to parameterize a resonance coupling to both channels using a pole interfering with a polynomial in an $S$-wave $K$-matrix,
\begin{equation}
	K_{\alpha\beta}(s) = \frac{g_\alpha g_\beta}{M^2 - s} + \gamma^{(0)}_{\alpha\beta} + \gamma^{(1)}_{\alpha\beta} \, s + \ldots, \label{Kmatrix}
\end{equation}
where the inverse of the $t^{\text{Res}}$-matrix   is   given by 
\begin{equation}
	\left[ \left (t^{\text{Res}} \right )^{-1}(s)\right]_{\alpha\beta} = \left[K^{-1}(s) \right]_{\alpha\beta} + \delta_{\alpha\beta}\, I_\alpha(s).
\end{equation}
Here $I_\alpha(s)$ is the Chew-Mandelstam form  \cite{Basdevant:1977} whose imaginary part above threshold ($s> 4m_\alpha^2$) is the phase-space,
\begin{equation}
	I_\alpha(s) = I_\alpha(0) - \frac{s}{\pi} \int_{4m_\alpha^2}^\infty \!\!\!\! ds' \sqrt{1-\frac{4m_\alpha^2}{s'} } \frac{1}{(s'-s)s'}.
\end{equation}
We have opted to subtract the integral once, and it is convenient to choose $I_\alpha(0)$ such that $\text{Re}\, I_\alpha(M^2) = 0$ so that we have an amplitude which for real $s$ near $M^2$ is close to the Breit-Wigner form with mass $M$.

 Given an explicit model for the scattering amplitudes, we can solve Eq.(\ref{latcoupcondition})  for the finite volume spectra in various volumes and total momenta $P= \frac{2\pi}{L}d, d\in \mathbb{Z}$.

\begin{widetext}
\begin{eqnarray}\label{latcoupcondition}
\left (\frac{1}{f^{\phi \phi}_{\mathcal{P}}} +i + \cot \frac{  p_{\phi} L + \pi d}{2} \right ) \left (\frac{1}{f^{\sigma \sigma}_{\mathcal{P}}} +i + \cot \frac{  p_{\sigma} L + \pi d}{2}  \right )= \left (i + \cot \frac{  p_{\phi} L + \pi d}{2} \right ) \left (i + \cot \frac{  p_{\sigma} L + \pi d}{2} \right ) \frac{ \left (f^{\phi \sigma}_{\mathcal{P}} \right  )^{2}}{f^{\phi \phi}_{\mathcal{P}} f^{\sigma \sigma}_{\mathcal{P}}} .\nonumber \\
\end{eqnarray}
\end{widetext}
 Eq.(\ref{latcoupcondition}) is derived from Eq.(\ref{coupcondition}) by replacing $ \gamma k_{\alpha}$ with $p_{\alpha}$ ($\alpha = \phi, \sigma$), where $p_{\alpha}$ is the relative momentum of two particles  in a moving frame: $p_{\alpha} = \frac{p_{\alpha,1}-p_{\alpha,2}}{2}$  and $P = p_{\alpha, 1}+p_{\alpha,2} $. To compensate for the ultraviolet cut-off effect from the finite  lattice spacing, the dispersion relation $\cosh E = \cosh m +1 - \cosh p$ \cite{Gatteringer:1993} is used in this work, accordingly, in Eq.(\ref{latcoupcondition}),    the relative momenta of two particles  in a moving frame $p_{\alpha}  $ are solved by the   equations    
\begin{eqnarray} 
P &=& p_{\alpha, 1}+p_{\alpha,2} , \nonumber \\
E^{(d)} &=& \sum_{i=1,2} \cosh^{-1} \left ( \cosh m_{\alpha} +1 -\cos p_{\alpha, i}  \right ). \nonumber
\end{eqnarray}
The lattice dispersion relation and finite size effects   are further discussed in Appendix \ref{latdispersion}.

     \begin{figure}
\begin{center}
\includegraphics[width=0.54\textwidth]{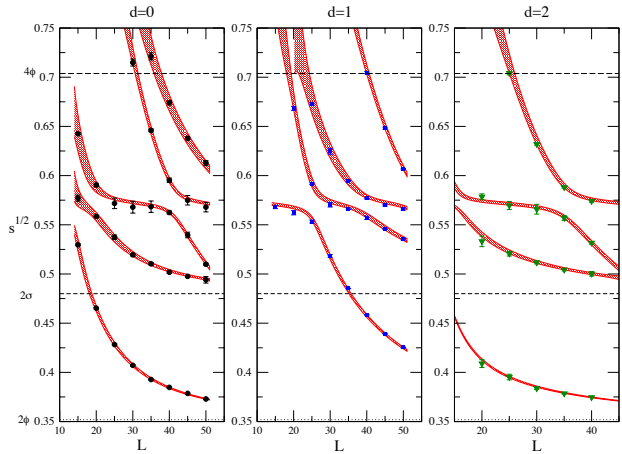}  
\caption{ The finite volume energy spectra from $K$-matrix model (red band)   as function of $(L,d)$, the spectra of $K$-matrix model are obtained  by performing the fit on $d=0$ lattice data below $4\phi$ threshold  only (black filled circles on the left). All the spectra in above three plots ($d=0,1,2$) are presented in CM frame.  The prediction of energy spectra from $K$-matrix model fit (red band) are also given for  (Middle) $d=1$ and (Right) $d=2$ compared to the  Monte Carlo  simulation data for  $d=1$ (blue filled square) and $d=2$ (green filled triangle) respectively.  $\frac{\chi^{2}}{N_{dof}} = \frac{20.4}{31-9}=0.93$. \label{speckmat}}
\end{center}
\end{figure}

   \begin{figure}
\begin{center}
\includegraphics[width=0.54\textwidth]{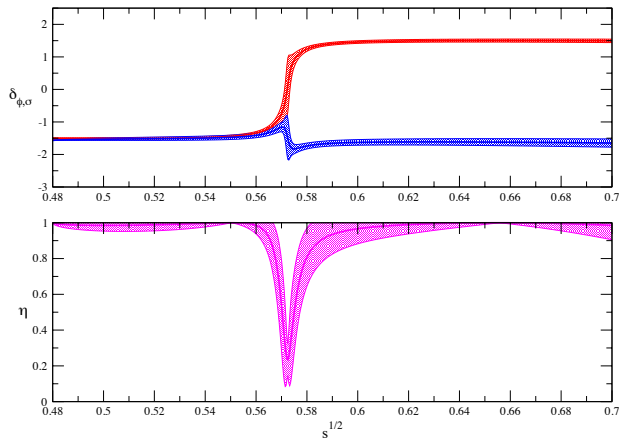}  
\caption{ The extracted phase shifts $\delta_{\phi}$(red), $\delta_{\sigma}$(blue) and inelasticity $\eta$(purple). \label{phaseinelast}}
\end{center}
\end{figure}

\subsection{Data analysis}
With the $K$-matrix model described in Section \ref{kmat}, we  can perform a global fitting method proposed in  \cite{Guo:2013cp}  to the spectra in Fig.\ref{spec}. For this purpose, we  can  minimize a function
\begin{equation}
\chi^2\big(\{a_i\}\big) = \sum_{E_n(L,d)} 
\frac{ \left[ E_n(L,d) - E^\text{det}_n(L,d; \{a_i\}) \right]^2
}{\sigma(E_n(L,d))^2}, \label{chisq}
\end{equation}
within the space of $K$-matrix parameters, $\{a_i\} = \{M, g_\phi, g_\sigma, \gamma^{(n)} \ldots \} $, where $E_n$ denotes the energy levels from Monte Carlo  simulation,  and $E^\text{det}_n $ are the solutions of  Eq.(\ref{latcoupcondition}).

Instead of  establishing the resonance pole position  in our toy model,  the purpose of this work is  to demonstrate (1) the methodology of extracting scattering amplitudes from  data of coupled-channel Monte Carlo  simulations,  (2) predictability of scattering amplitudes extracted from a set of lattice data,  and (3) the validity of our formalism  while taking into account of  the   finite size effect  presented in Appendix \ref{latdispersion}.  Therefore, in this work,  we choose to  fit the spectra below $4\phi $ threshold for $d=0$ only, then for a consistency check, we compare our predicted spectra for $d=1,2$ to the spectra from the Monte Carlo  simulation.   We show the spectra of $K$-matrix model (red bands) in  Fig.\ref{speckmat} with the comparison of spectra from the Monte Carlo  simulation (filled black circles, filled blue squares and filled green triangles).  The $K$-matrix we used in the fitting has   nine free parameters,  the polynomial of $\gamma^{(n)}$ is taken up to $\mathcal{O}(s^{1}) (n=0,1)$. The value of parameters we find from   fitting read
\begin{eqnarray}
M = 0.572 (1)   ,   g_{\phi} = 0.064(4),  g_{\sigma} = 0.060(4), \nonumber \\
\gamma^{(0)}_{\phi \phi} = 0.3(1) ,  \gamma^{(1)}_{\phi \phi} =   -0.7(3)  ,  
\gamma^{(0)}_{\phi \sigma} =0.11(3)    ,   \nonumber \\
\gamma^{(1)}_{\phi \sigma} = - 0.3(1)     , \gamma^{(0)}_{\sigma \sigma} =-0.6(2)    ,  \gamma^{(1)}_{\sigma \sigma} =  1.5(5)  . \nonumber
\end{eqnarray}
The extracted phase shifts $\delta_{\phi}$, $\delta_{\sigma}$ and inelasticity $\eta$ are shown in Fig.\ref{phaseinelast}.

As  demonstrated in the middle and the right plots in Fig.\ref{speckmat},   our predicted energy spectra from  $K$-matrix model (red bands) for $d=1,2$  agree with the spectra from the Monte Carlo  simulation (filled blue squares and filled green triangles) within a reasonable precision. Therefore, we   accomplished our goals, (1) we proved that our formalism gives the consistent result in different moving frames while taking into account of finite size effect, and (2) we have shown that the global fitting method   is a valid and fairly reliable means for extracting scattering amplitudes from Monte Carlo  simulation data.

\section{Summary }\label{summary}

Based on the  Lippmann-Schwinger equation approach,  in Section \ref{singlechannel} and    \ref{coupledchannel},  we first derived   a generalized L\"uscher's formula in 2D   for two particles scattering in both the elastic and coupled-channel cases in moving frames. In Section  \ref{isingmodel},  we presented a 2D coupled-channel   scattering lattice model. The model   simulates a two-coupled-channel resonant  scattering   system, in which a resonance couples to both   channels.  Next, we performed   Monte Carlo simulations  on various finite lattice sizes and in different moving frames. The discrete finite-volume spectra were extracted by fitting two-particle correlation functions. 
 Finally, we used the 2D   generalized L\"uscher's formula   to extract the scattering amplitudes for the coupled-channel system from the discrete finite-volume spectra.  We have shown that  the global fitting method  can be  used to  reliably  extract scattering amplitudes from Monte Carlo  simulation data.   The finite size effects on the solution of the  generalized L\"uscher's formula were discussed  in details in Section \ref{isingmodel} and Appendix \ref{latdispersion}. We  demonstrated  that while taking into account of finite size effects,  our formulae  produce   consistent results in different moving frames.

\section{ACKNOWLEDGMENTS}
We thank  Dru~B.~Renner, Robert~G.~Edwards  and Han-Qing Zheng  for useful discussions, and the special thanks go to Dru~B.~Renner  for inspiration of this work and for his  encouragement. We also thank David~J.~Wilson for carefully reading through this manuscript.  PG acknowledges support from U.S. Department of Energy contract DE-AC05-06OR23177, under which Jefferson Science Associates, LLC, manages and operates Jefferson Laboratory.

\appendix

\section{One dimensional infinite sum }\label{expgreen}
Let's consider the one dimensional infinite sum in Eq.(\ref{greenpole}), 
\begin{equation}\label{infinitesum} 
 \sum_{n \in \mathbb{Z}}  e^{ i k |x  -x'- \gamma n L|} e^{i \frac{P}{2} n L}, 
\end{equation}
where $P=\frac{2\pi}{L}d, d \in \mathbb{Z}$.
\begin{widetext}
In the region which we are interested in:  $|x| > |x'|$ and $|x-x'| < \gamma L$,  Eq.(\ref{infinitesum}) can be rewritten to
\begin{equation} 
e^{ i k |x  |} \sum_{\mathcal{P}= \pm} Y_{\mathcal{P}} (x) Y_{\mathcal{P}} (x')   J^{*}_{\mathcal{P}} (k x')+ \sum^{n \neq 0}_{n \in \mathbb{Z}}  e^{ i k |  \gamma n L|} e^{i \frac{P}{2} n L}  \sum_{\mathcal{P}=\pm } Y_{\mathcal{P}} (n)  Y_{\mathcal{P}} (x) J_{\mathcal{P}} (k x)     \sum_{ \mathcal{P}' = \pm}Y_{\mathcal{P}'} (n) Y_{\mathcal{P}'} (x') J^{*}_{\mathcal{P}'} (k x').   
\end{equation}
With the help of equations
\begin{equation} 
 \sum^{n \neq 0}_{n \in \mathbb{Z}}  e^{ i k |  \gamma n L|} e^{i \frac{P}{2} n L} Y_{+}(n)= \frac{\cos \frac{PL}{2}  -e^{ i \gamma k L}}{ \cos  \gamma kL - \cos \frac{PL}{2}  }   ,     \ \  \ \ \ \  \sum^{n \neq 0}_{n \in \mathbb{Z}}  e^{ i k |  \gamma n L|} e^{i \frac{P}{2} n L}  Y_{-}(n)=i  \frac{\sin \frac{PL}{2}  }{ \cos  \gamma kL - \cos \frac{PL}{2}  }   ,    \nonumber
\end{equation}
where the infinite sums are performed by using  the property of polylogarithmic function  $Li_{0}(x) =\sum_{n=1}^{\infty} x^{n} = \frac{x}{1-x}$,  thus, we find
\begin{equation} 
 \sum_{n \in \mathbb{Z}}  e^{ i k |x  -x'- \gamma n L|} e^{ i \pi    n d } 
= \sum_{\mathcal{P} = \pm} Y_{\mathcal{P}} (x) Y_{\mathcal{P}} (x')  J^{*}_{\mathcal{P}} (k x')   \left [ e^{ i k |x  |} -  \left (1- i \cot \frac{\gamma k L + \pi d}{2} \right )    J_{\mathcal{P}} (k x) \right ], 
\end{equation}
for  $|x| > |x'|$ and $|x-x'| < \gamma L$.
\end{widetext}

  \begin{figure}
\begin{center}
\includegraphics[width=0.46\textwidth]{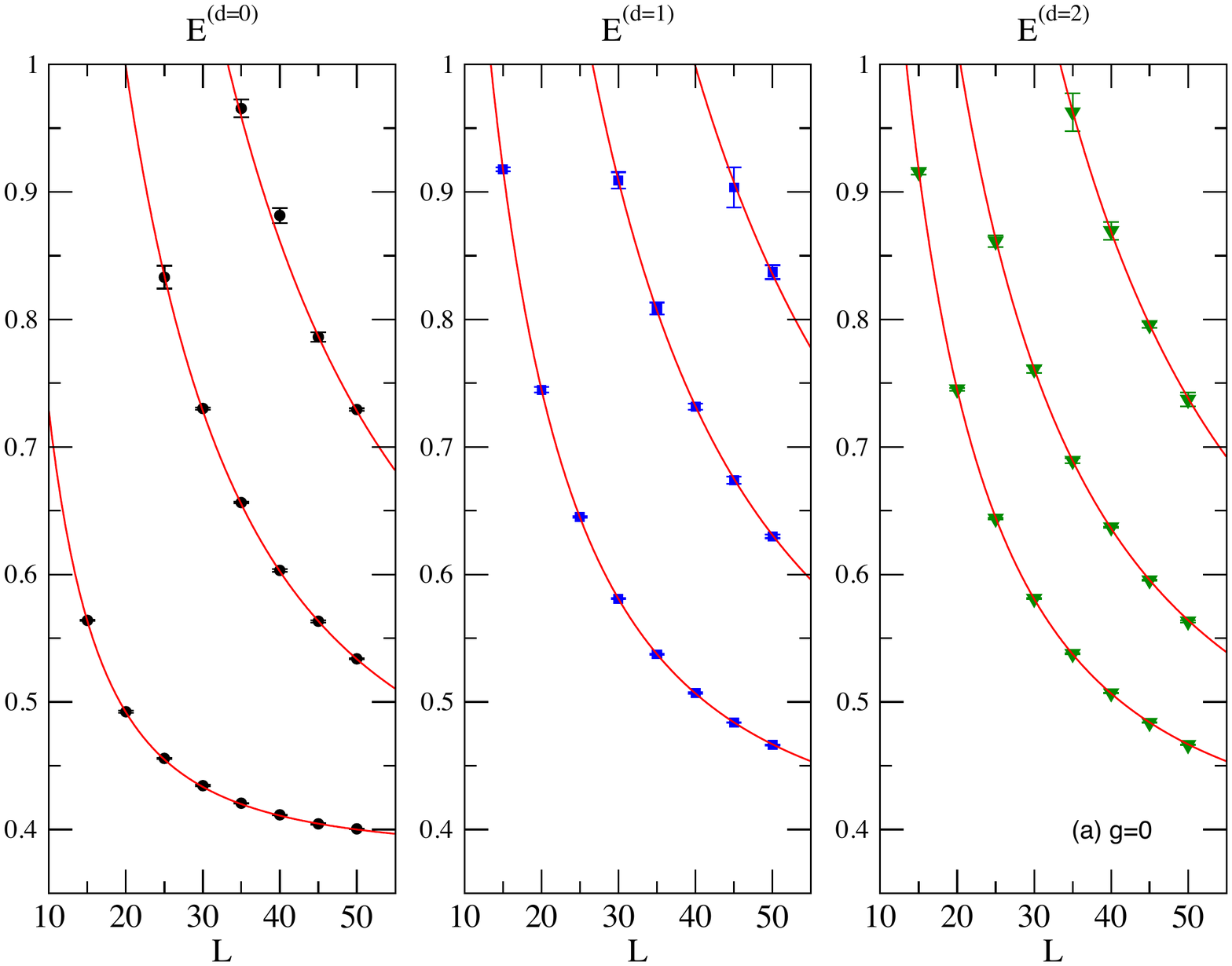}  
\includegraphics[width=0.46\textwidth]{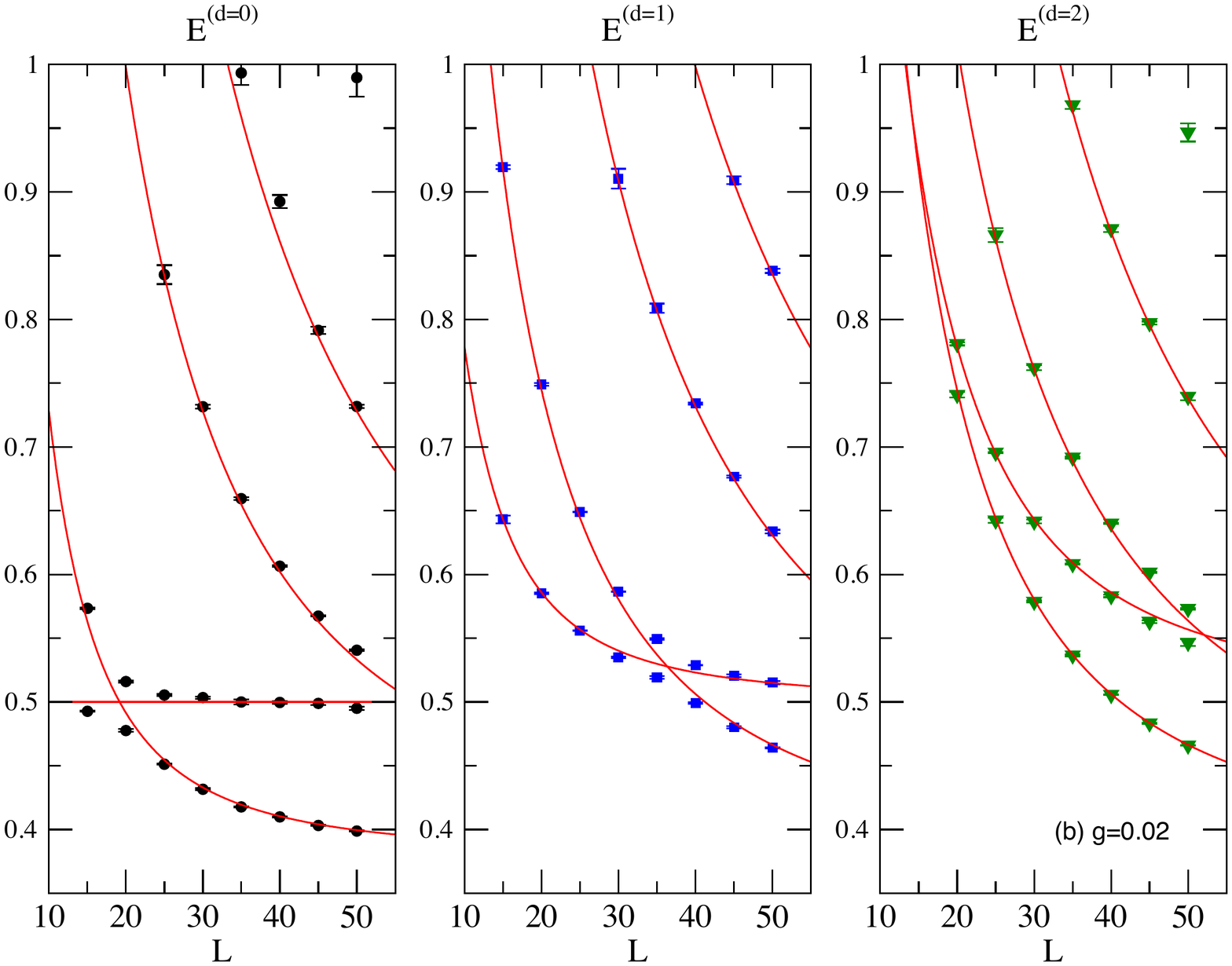} 
\includegraphics[width=0.46\textwidth]{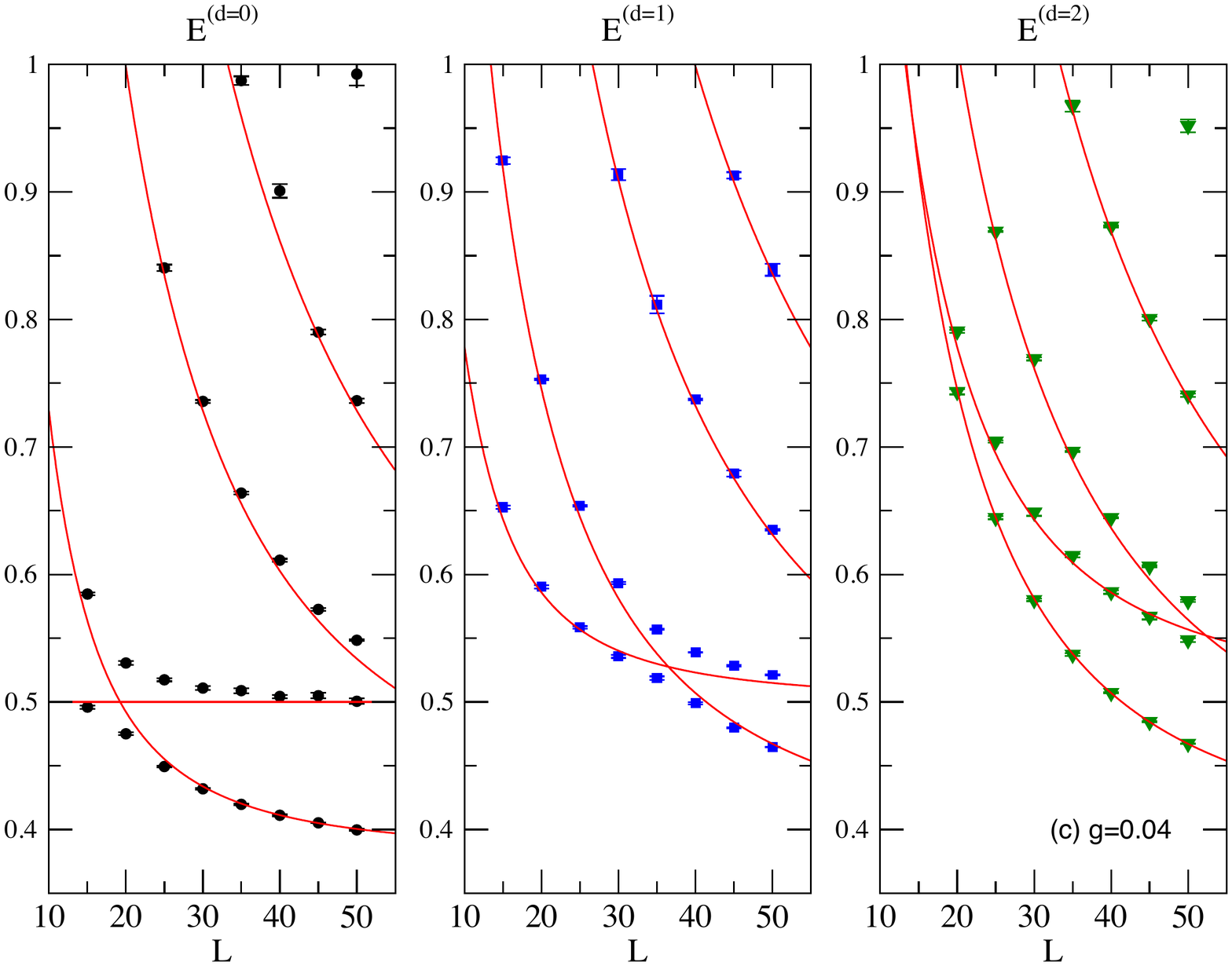} 
\caption{ The  energy spectra of single channel Ising model \cite{Gatteringer:1993} as function of $(L,d)$  for (a) $g=0$ (Upper panel), (b) $g=0.02$ (Middle panel) and (c)  $g=0.04$ (Lower panel)  respectively. In each panel, from left to right,  each individual plot is related to   $d=0$ (black  filled circles),  $1$ (red filled squares) and  $2$ (green  filled triangles) respectively.  The red curves represent (I) the energy spectra of a non-interacting pair of particles:  $E^{(d)} = \sum_{i=\pm} \cosh^{-1} \left ( \cosh m +1 -\cos  p_{i}  \right )$, where $p_{\pm }=\frac{2 \pi}{L} n_{\pm} \pm \frac{\pi}{L}$,  $n_{+}+n_{-} = d$ and    $(n_{\pm},  d) \in \mathbb{Z}$; (II) the    energy spectra of a stable resonance: $E^{(d)} =  \cosh^{-1} \left ( \cosh m_{\rho} +1 -\cos  P  \right )$, where $m_{\rho} \simeq 0.5$. \label{specsingle}}
\end{center}
\end{figure}

  \begin{figure}
\begin{center}
\includegraphics[width=0.54\textwidth]{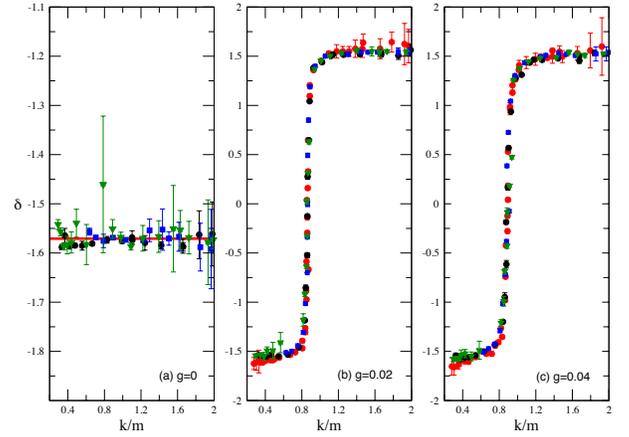}  
\caption{ The  phase shifts of single channel Ising model \cite{Gatteringer:1993}   for (a) $g=0$ (Left), (b) $g=0.02$ (Middle) and (c)  $g=0.04$ (Right)  respectively. The red line in left plot labels the theoretical expectation $\delta_{\text{Ising}} = -\frac{\pi}{2}$   for non-interacting case (a) $ g=0$.  The red  filled  circles, black  filled  circles, blue filled  squares and green filled triangles represent  the    result in   Fig.6 of \cite{Gatteringer:1993},  and our results  for $d=0,1,2$  respectively. \label{deltasingle}}.
\end{center}
\end{figure}

\section{Lattice dispersion relation}\label{latdispersion}
For the  determination of the lattice spectrum at a precise level, the finite size effect  has to be taken into consideration by  using the lattice dispersion relation \cite{Gatteringer:1993}
\begin{equation}\label{latdisp}
\cosh \sqrt{s} = \cosh E^{(d) } - (1- \cos P), 
\end{equation} 
where $E^{(d )}$  and $ \sqrt{s}$ are the total energy of system in moving frames and the CM frame respectively,  and  the total momentum of system is given by $P=\frac{2\pi}{L}d, d \in \mathbb{Z}$.  In the limit of vanishing lattice spacing,  Eq.(\ref{latdisp}) reduces to the relativistic dispersion relation: $E^{(d) }    =\sqrt{ s + P^{2} }$. 

The relative momentum of two particles $p=\frac{p_{1}-p_{2}}{2}$ in a moving frame ($p_{1}+p_{2}=P $) is related to the relative momentum of two particles $k= \frac{k_{1}-k_{2}}{2}$ in the CM frame ($k_{1}+k_{2}=0$) by Lorentz transformation relation $p =\gamma k$.  In the limit of vanishing lattice spacing, the Lorentz contraction factor $\gamma$ is given by $\gamma= \frac{E^{(d)}}{\sqrt{s}}$. However, due to the ultraviolet cut-off effect from finite  lattice spacing, the definition of the Lorentz contraction factor $\gamma= \frac{E^{(d)}}{\sqrt{s}}$  is inconsistent with lattice dispersion relation Eq.(\ref{latdisp}). This inconsistency leads to the large discrepancies of phase shifts and inelasiticy computed in different frames.
To resolve this problem, we may use the relation $p =\gamma k$ to rewrite Eq.(\ref{singeq}),  Eq.(\ref{coupcondition}) and Eq.(\ref{coupeq}) to 
\begin{equation}\label{latsingeq} 
\cot \delta_{\mathcal{P}}+ \cot \frac{ p L + \pi d}{2}   =0 , 
\end{equation}
for single channel scattering, and 
\begin{widetext}
\begin{equation} 
\left (\frac{1}{f^{\phi \phi}_{\mathcal{P}}} +i + \cot \frac{  p_{\phi} L + \pi d}{2} \right ) \left (\frac{1}{f^{\sigma \sigma}_{\mathcal{P}}} +i + \cot \frac{  p_{\sigma} L + \pi d}{2}  \right )= \left (i + \cot \frac{  p_{\phi} L + \pi d}{2} \right ) \left (i + \cot \frac{  p_{\sigma} L + \pi d}{2} \right ) \frac{ \left (f^{\phi \sigma}_{\mathcal{P}} \right  )^{2}}{f^{\phi \phi}_{\mathcal{P}} f^{\sigma \sigma}_{\mathcal{P}}} .\nonumber 
\end{equation}
\end{widetext}
or
 \begin{equation}\label{latcoupeq}  
\eta_{\mathcal{P}} \left (-1 \right )^{d} = \frac{\cos \left (   \frac{ p_{\phi} + p_{\sigma}}{2} L+ \delta^{\phi}_{\mathcal{P}} + \delta^{\sigma}_{\mathcal{P} } \right )}{\cos \left (   \frac{ p_{\phi} - p_{\sigma}}{2} L + \delta^{\phi}_{\mathcal{P}} - \delta^{\sigma}_{\mathcal{P} } \right )},
\end{equation}
for coupled channel scattering, respectively. In Eq.(\ref{latsingeq}), Eq.(\ref{latcoupcondition})  and Eq.(\ref{latcoupeq}), the relative momentum of two particles  $p = \frac{p_{1}-p_{2}}{2}$ is    solved by   equations    
\begin{eqnarray}\label{lattotE}
P&=&p_{1}+p_{2} , \nonumber \\
E^{(d)} &=& \sum_{i=1,2} \cosh^{-1} \left ( \cosh m +1 -\cos p_{i}  \right ).
\end{eqnarray}
So that, rather than solving  Eq.(\ref{singeq}),    Eq.(\ref{coupcondition})  and   Eq.(\ref{coupeq})    with the Lorentz contraction factor given by $\gamma= \frac{E^{(d)}}{\sqrt{s}}$,  we use  Eq.(\ref{latsingeq})  and Eq.(\ref{latcoupcondition})   with the solution of relative momentum of two particles given by Eq.(\ref{lattotE})     for single and coupled-channel scattering respectively  in this work.

As a simple demonstration how the above proposal works,   let's consider  a  non-interacting  two-particle system in $1+1$ dimensions.   Two  particles  in an arbitrary  moving frame, $P= \frac{2\pi}{L}d$, have individual  momenta  $p_{\pm}=\frac{2 \pi}{L} n_{\pm} \pm \frac{\pi}{L}$ respectively,    where $n_{+}+n_{-} = d$ and  $(n_{\pm},  d) \in \mathbb{Z}$. So that we get a relation:  $\frac{pL + \pi d}{2} =n_{+} \pi + \frac{\pi}{2}$, where $p =\frac{p_{+}-p_{-}}{2}$ is the relative momentum of two-particle system. Using Eq.(\ref{latsingeq}),   we conclude that  the   phase shift of  two non-interacting particles is given by $\delta = \delta_{\text{Ising}} = -\frac{\pi}{2}$. This conclusion  derived from  Eq.(\ref{latsingeq})   holds in all the moving frames. 

As an more  quantitative   example, we generalize  the single channel Ising model  computed in CM frame   in \cite{Gatteringer:1993}  to the moving frames, we compute the three sets of models \cite{Gatteringer:1993}  (see Table 1 in  \cite{Gatteringer:1993})   for three different moving frames:  $d=0,1,2$.  Three models are  labeled by   coupling constants:  (a) $g=0$, (b) $g=0.02$ and (c) $g=0.04$, all the parameters are given by Table 1 in  \cite{Gatteringer:1993}.  The measured energy levels for three models ($g=0,0.02,0.04$) and three moving frames ($d=0,1,2$) are shown in  Fig.\ref{specsingle}, the extracted phase shifts by using Eq.(\ref{latsingeq}) along with the solution of relative momentum of two particles from  Eq.(\ref{lattotE})  are shown in Fig.\ref{deltasingle}.  Fig.\ref{deltasingle} demonstrates the consistent calculation of phase shift from different moving frames by using Eq.(\ref{latsingeq}) along with  Eq.(\ref{lattotE}).

\end{document}